\documentstyle[aps,prl,twocolumn]{revtex}

\begin{document}

\draft
\title{Comment on ``Determination of Interlayer Diffusion Parameters
for Ag/Ag(111)''}
\author{Joachim Krug}
\address{
Fachbereich Physik, Universit\"at Essen, 45117 Essen, Germany}

\date{\today}

\maketitle

% \pacs{68.35.Fx, 61.14.Hg, 61.16.Ch, 68.55.-a}

In a recent Letter \cite{roos} Roos and Tringides (RT) collected 
evidence in support of the
hypothesis that the preexponential factor $\nu_s$
in the interlayer jump rate on Ag(111) exceeds the corresponding 
quantity $\nu_t$
for inlayer transport by about two orders of magnitude.
As part of the argument they present a simplified analysis of an 
experiment carried out by Bromann {\it et al.} (BBRK)
\cite{brune}, in which second layer nucleation on top of
predeposited Ag islands on Ag(111)
was investigated. Here I point out a mistake in the analysis of RT, 
and show that the correct application of their idea does not allow
to conclude that $\nu_s/\nu_t > 1$. I then discuss more broadly our
present understanding of interlayer diffusion on Ag(111)
in view of a recent reanalysis \cite{krug00} of the BBRK experiment.

RT estimate the probability $f$ that an atom deposited during the second
dose descends from the island within the time $\tau$ between successive
deposition events.
They write $f = (\lambda/d) p$,
where $\lambda = \sqrt{D_t \tau}$ is the diffusion length, 
$d$ the island diameter and $p$ the probability that an atom poised
at an edge site jumps down from the island. The ratio 
$\lambda/d$ is referred to by RT as ``{\it
the number of edge interrogations}''.
In fact this number is much larger 
than $\lambda/d$. Provided that $\lambda \gg d$, which is the
case of interest here, it is given by the number of
diffusion jumps $D_t \tau = \lambda^2$ multiplied by
the fraction $L/A$ of edge sites among all sites on the island; 
$L$ is the island perimeter and $A \sim d^2$ the island area in 
units of the lattice constant, so 
$(L/A) D_t \tau \sim L (\lambda/d)^2 \gg \lambda/d$.

To obtain the correct expression for $f$, note that
the probability that an atom on the island descends
in a small time interval $dt$ is 
$(L/A) D_s (1-f) dt$, 
where $D_s = p D_t = (\nu_s/\nu_t) D_t 
\exp(-\Delta E_s/k T)$ is the 
interlayer hopping rate.
Integrating up to time $t = \tau$ then yields 
\begin{equation}
\label{f}
f = 1 - \exp[-(L/A) D_s \tau],
\end{equation}
which corrects Eq.(1) of RT.
Using the numbers given by RT, for 
the case of islands of radius 30 ${\rm \AA}$ at temperature
$T = 130$ K, we obtain $f = 1 - \exp[-130 (\nu_s/\nu_t)]$
for $\Delta E_s = 0.13$ eV. Thus $f = 1$ provided $\nu_s/\nu_t 
\geq 0.1$.

In addition to the BBRK experiment, RT base their conclusion on
the analysis of two other growth situations involving interlayer transport.
In all three cases they work at a single temperature, which implies
that information about $\nu_s$ can be extracted only if $\Delta E_s$
is known. The value of $\Delta E_s$ used by RT was obtained
in \cite{brune} by analyzing
the dependence of the fraction of islands with second layer
nuclei (a quantity somewhat similar to $1-f$) as a function of the radius
of predeposited islands, at two different temperatures
$T = 120$ K and 130 K. The analysis was based on 
a theory of second layer nucleation 
due to Tersoff {\it et al.} \cite{tersoff}, which has 
recently been shown to be quantitatively incorrect
\cite{krug00,maass,krug00b}.
Other groups \cite{others} have estimated $\Delta E_s$ assuming that 
the adatom density at second layer nucleation is comparable
to that at which first layer islands nucleate, which is generally
not true \cite{krug00}. 

A reanalysis \cite{krug00} of the BBRK data using the correct expression
for the rate of second layer nucleation yields  
a very large step edge
barrier $\Delta E_s \approx 0.32$ eV, 
accompanied by a prefactor $\nu_s  \approx 8 \times 10^{19}$ s$^{-1}$,
which would imply $\nu_s/\nu_t \approx 4 \times 10^8$. 
As such a large preexponential
factor seems hard to justify physically, this suggests that 
a single pair of diffusion parameters is insufficient to describe
interlayer transport on Ag(111), and that additional 
second layer nucleation experiments
at variable temperature are called for. 
A temperature-dependent measurement of interlayer transport
based on the decay of vacancy islands
was performed by Morgenstern {\it et al.}, who estimate
$\Delta E_s = 0.13$ eV and $\nu_s/\nu_t = 10^{-0.6 \pm 0.5} < 1$
\cite{morgenstern}. The discrepancy between these numbers and
the (correctly analyzed) BBRK experiment indicates that, despite
considerable effort, interlayer diffusion for Ag(111)
remains an open problem.  

Useful correspondence with Michael Tringides is gratefully acknowledged. 
This work was 
supported by DFG within SFB 237.

\end{document}